\documentclass[fleqn,twoside]{article}
\usepackage{gc,amsfonts,cite}

\eqsection

\heads
{I.S. Goncharenko, V.D. Ivashchuk, S. Rudametkin-Aguilar and D. Singleton}
{Electric S-brane solutions with a parallel charge density form
	on Ricci-flat factor space}

\begin{document}
\twocolumn[
\prepno{gr-qc/0609114}{\GC{12} 169 (2006)}
%%\AH {169}{172}

\Title {ELECTRIC S-BRANE SOLUTIONS WITH A PARALLEL CHARGE \yy
	DENSITY FORM ON A RICCI-FLAT FACTOR SPACE}

\Authors{I.S. Goncharenko, V.D. Ivashchuk\foom 1}{, S. Rudametkin-Aguilar
	and D.  Singleton\foom2}
{Centre for Gravitation and Fundamental Metrology, VNIIMS, and\\
 Institute of Gravitation and Cosmology, Peoples' Friendship
  	University of Russia,\\
 6 Miklukho-Maklay St., 117198 Moscow, Russia}
{Physics Department, CSU Fresno, Fresno, CA 93740-8031, USA  }

%\Rec{21 May 2006}

\Abstract
{We generalize the previously studied cosmological solutions in
$D$-dimensional gravity with an antisymmetric $(p+2)$-form to the case when
the spatial part of the metric is Ricci-flat rather than flat. These
generalized solutions are characterized by a parallel self-dual or
anti-self-dual charge density form $Q$ of rank $2 m$ and satisfy the
condition $Q^2 >0$. As with the previous flat-space case, these electric,
S-brane solutions only exist when  $D=4m+1 = 5, 9, 13, ...$ and  $p= 2m -1 =
1, 3, 5, ...$. We further generalize these solutions by adding
Ricci-flat factor-spaces which are not covered by branes. }

%\RAbstract{Электрические S-бранные решения с параллельной формой плотности
%заряда на риччи-плоском фактор-пространстве}
%{И.С. Гончаренко, В.Д. Иващук, С. Рудаметкин-Агилар, Д. Синглтон}
%{Ранее изучавшиеся решения в $D$-мерной гравитации с антисимметричной
%$(p+2)$-формой обобщаются на случай, когда пространственная часть метрики
%риччи-плоская. Эти обобщенные решения характеризуются самодуальной или
%антисамодуальной формой зарядовой плотности $Q$ ранга $2m$ и 
%удовлетворяют условию $Q^2 > 0$. Как и в предыдущем 
%пространственно-плоском случае, электрические $S$-бранные решения 
%появляются только когда размерность $D= 4m +1=5,9,13,\dots$ и 
%$p=2m-1=1,3,5,\dots$. Получено дальнейшее обобщение этих решений 
%добавлением дополнительных риччи-плоских фактор-пространств, не покрытых 
%бранами. }

 ]       %%%%%%%%%%%%%%%%%%%%
 \email 1 {rusgs@phys.msu.ru}
 \email 2 {dougs@csufresno.edu}

\section{\bf Introduction}

In the paper, we continue the study of $S$-brane solutions with a maximum
number of electric branes \cite{IS,IMS,DIM}. $S$-branes are brane solutions
having ``warp'' factors which are time-dependent rather than spatially
dependent. They are thought to play a role in higher-dimensional
cosmological models, especially string theory inspired cosmologies. A
sampling of the work done on $S$-brane solutions and their physical
applications can be found in [4--18] and references therein.

In \cite{IS}, electric $S$-brane solutions, with a maximum number of
orthogonal branes, were constructed. These solutions only occurred for
$D=4m+1$ spacetime dimensions with $(2m +1)$-forms.  They also had the
property that the charge density forms of the electric branes were either
self-dual or anti-self-dual. For example, in the case when $D=5$ with
$3$-forms, the six electric branes obeyed the following relations:
\bear \label{1.1}
	Q_{12} = \mp Q_{34}, \quad\ Q_{13} = \pm Q_{24}, \quad\
     	Q_{14} = \mp Q_{23}.
\ear
or, equivalently,
\beq \label{b2}
      Q_{i j} =  \pm \frac{1}{2} \eps _{i j k l}
      Q^{k l} = \pm (* Q)_{i j}.
\eeq
The solution (metric plus form fields and scalar fields) for the case when
all the $Q_{i j}$, $i < j$, were non-zero, followed from the restrictions
coming from the non-diagonal part of Hilbert-Einstein equations.

Some of the cosmological consequences of these maximal electric $S$-brane
solutions (in case of the absence of scalar fields) were investigated in
\cite{IMS}, e.g., suppression of oscillating behaviour near the singularity
due to constraints coming from the diagonal form of the metric was found.
(This is important for the billiard approach for models with branes
suggested in \cite{IMb1}).

In addition to their relation to cosmology, the study of electric S-branes
with various charge densities may be of possible interest to string theory
because of the important role played by the relations between the charge
densities of $D$-branes \cite{Polchinsky}. Such relations in the general
case may be described mathematically by $K$-theory (see \cite{MM,Witten} and
references therein). Here, as in our previous work \cite{IS,IMS}, the
relationships between the charge densities follow directly from the
equations of motion.

\section{S-brane solution with flat factor space}

\subsection{$D$ dimensional gravity and $(p+2)$-form}

As in \cite{IS}, we consider the model governed by the action
\beq\label{2.1i}
    S = \int_{M} d^{D}z \sqrt{|g|} \left[ {R}[g]
    		-  \frac{1}{q!}  F^2 \right],
\eeq
where $g = g_{MN} dz^{M} \otimes dz^{N}$ is the metric, ${R}[g]$ is
the $D$-dimensional Ricci scalar calculated from the metric $g$, and
\[
   F =  dA = \frac{1}{q!} F_{M_1 \ldots M_{q}}
   dz^{M_1} \wedge \ldots \wedge dz^{M_{q}}
\]
is a $q$-form, $q =  p +2 \geq 1$. For $p=0$ one would have a 2-form
analogous to the Maxwell field strength tensor. Thus the action considered
in (\ref{2.1i}) gives $D$-dimensional gravity coupled to an
``electromagnetic'' field. In (\ref{2.1i}), we denote $|g| = |\det(g_{MN})|$
and  $F^2 = F_{M_1 \ldots M_{q}} F_{N_1 \ldots N_{q}}
         g^{M_1 N_1} \ldots g^{M_{q} N_{q}}$,
The equations of motion corresponding to  (\ref{2.1i}) are
\bear\label{2.4i}
	   R^M_N - \frac{1}{2} \delta^M_N R  \eql   T^M_N,
\\   \label{2.6i}
     		\nabla_{M_1}[g] F^{M_1 \ldots M_{q}}  \eql  0.
\ear
In (\ref{2.6i}), ${\nabla}[g]$ is the covariant derivative operator
corresponding to $g$. \eqs (\ref{2.4i}) and (\ref{2.6i}) are the
multidimensional Einstein-Hilbert equations and the ``Maxwell'' equations
for the $q$-form, respectively. The source term in (\ref{2.4i}),
the stress-energy tensor of  $q$-form, reads
\bearr\label{2.9i}
       T^M_N [F,g] = \frac{1}{q!} \left[ - \frac{1}{2} \delta^M_N F^2
   	+ q  F_{N M_2 \ldots M_{q}} F^{M M_2 \ldots M_q}\right] .
\nnn
\ear

\subsection{Cosmological type solutions for  $ D=5, 9, 13, ...$}

Here we give a quick summary of $Sp$-brane solutions from \cite{IMS}
in dimensions
\beq\label{3.h}
         D= n+1 = 4m + 1= 5, 9, 13, \dots
\eeq
with $(p+2)$-forms whose rank is given by
\beq\label{3.p}
         p = 2m - 1= 1, 3, 5, \dots .
\eeq
These solutions are defined on  the manifold
\beq\label{2.10g}
    	M = (t_{-}, t_{+})  \times \mathbb{R}^{n}
\eeq
and have the following form:
\bear\label{4.pa}
      ds^2 \eql - \e^{2n \phi(t)} dt^2 + \e^{2 \phi(t)}
     			\sum _{i=1}^n (dy^i)^2,
\\     \label{4.pc}
     F \eql \e^{2 f(t)} dt \wedge Q,
\\          \label{4.pq}
      Q \eql  \frac{1}{(p+1)!} Q_{i_0  \dots i_p}
                           dy^{i_0}  \wedge \dots  \wedge dy^{i_p}.
\ear
$Q_{i_0  \dots i_p}$ are constant components of the charge density form $Q$.
$Q$ is self-dual or anti-self-dual in flat Euclidean space $\mathbb{R}^n$,
i.e.,
\bearr\label{3.eaa}
      Q_{i_0  \dots i_p} =
       \pm \frac{1}{(p+1)!} \eps _{i_0  \dots i_p j_0 \dots j_p}
       Q^{j_0 \dots j_p} = \pm (* Q)_{i_0  \dots i_p}.
\nnn
\ear
We now put
\beq\label{4.q}
        Q^2 \equiv \frac{1}{(p+ 1)!}
           \sum_{i_0,  \dots, i_p} Q_{i_0  \dots i_p}^2 > 0,
\eeq
i.e., at least one charge density is non-zero: $Q_{i_0 \dots i_p}
\neq 0$, for some $i_0 <  \dots < i_p$. Then it can be shown \cite{IS} that
\eqs (\ref{2.4i}) (\ref{2.6i}) reduce to the set of equations
\beq\label{4.f}
  	{\ddot f} = {\dot f}^2 = -Q^2 K \e^{2 f}.
\eeq
The dots denote derivatives with respect to $t$ and
\beq\label{4.k}
          K \equiv   - \frac{n}{4(n-1)}.
\eeq
The function $\phi (t)$ is given in terms of $f(t)$ by
\beq\label{4.n}
        \phi(t) = \frac{2}{n} f(t).
\eeq
Thus, the solution of the system is determined once (\ref{4.f}) is
solved. A solution for $f(t)$ is given by
\beq\label{4.i}
        f = - \ln \left[|t - t_0 ||K Q^2|^{1/2} \right].
\eeq
The above solution describes a collection of $l \leq \frac{(4m)!}{(2m)!^2}$
electric $Sp$-branes.

  \section{\bf Generalization to Ricci-flat factor space}

Here we generalize the solution from the previous section to the case when
the manifold (\ref{2.10g}) is replaced by the manifold
\beq\label{3.10g}
       M = (t_{-}, t_{+})  \times N,
\eeq
where $N$ is an $n$-dimensional oriented  manifold of dimension $n =4m$, $m
=1,2, \dots$, equipped with the Ricci-flat metric $h = h_{ij}(y)dy^i \otimes
dy^j$ of Euclidean signature,  i.e., the Ricci tensor calculated from $h$ is
zero, $R_{ij}[h] =0$. Let
 \beq\label{3.11}
      Q =  \frac{1}{(p+1)!} Q_{i_0  \dots i_p}(y)
              dy^{i_0}  \wedge \dots  \wedge dy^{i_p}
 \eeq
be a form of rank $2m$ defined on the manifold $N$. The components
$Q_{i_0 \dots i_p} (y)$ now depend on the spatial coordinates $y^i$. The
form $Q$ satisfies two requirements. First, it is parallel, i.e.,
covariantly constant w.r.t. $h$:
\beq\label{3.12a}
      {\rm (i)} \cm \nabla [h] Q = 0,
\eeq
and, second, it is self-dual or anti-self-dual:
\beq\label{3.12b}
      {\rm (ii)} \cm	Q = \pm * Q.
  \eeq
Here $* = *[h]$ is the Hodge operator corresponding to the metric $h$.
From condition (i) one finds that
\beq\label{3.12ca}
       Q^2 \equiv \frac{1}{(p+ 1)!} h^{i_0 j_0} \dots h^{i_p j_p}
       		Q_{i_0  \dots i_p} Q_{j_0  \dots j_p}
  \eeq
is  constant. It also follows from (\ref{3.12a}) that $Q$ is closed:
\beq\label{5.b}
        d Q =  0.
\eeq
In what follows we put
\beq\label{3.12c}
          Q^2 > 0,
\eeq
or, equivalently, $Q$ is non-zero.

For the metric and $p$-form, we take the following ansatz on the manifold
(\ref{3.10g}) similar to the ansatz used in the flat case, (\ref{4.pa}),
(\ref{4.pc}):
\bear\label{5.pa}
      ds^2  \eql - \e^{2n \phi(t)} dt^2 + \e^{2 \phi(t)}
      				h_{ij} (y) dy^i dy^j,
\\         \label{5.pc}
       F \eql \e^{2 f(t)} dt \wedge Q.
\ear
The $2m$-form $Q$ is defined in (\ref{3.11}) and satisfies (\ref{3.12a}),
(\ref{3.12b}) and (\ref{3.12c}). Except for the spatial metric $h$, this is
similar to the previous flat case ansatz. In fact, we will find that the
form of these new (``Ricci-flat'') solutions is essentially the same as in
the flat case. $K$ and the function $\phi (t)$ are again given by
(\ref{4.k}) and (\ref{4.n}). As before, once $f(t)$ is given, the entire
solution is obtained. The function $f(t)$ again obeys \eq (\ref{4.f}), and
thus the solution is (\ref{4.i}).

Now we show that the metric with the line element (\ref{5.pa}) and the
$(2m+1)$-form (\ref{5.pc}) do satisfy the field equations (\ref{2.4i}),
(\ref{2.6i}) for the case when the manifold is given by (\ref{3.10g}).
Because of (\ref{3.12a}), the ``Maxwell'' equations (\ref{2.6i}) reduce to
those for the flat-case solution from the previous section, and so the
$(p+2)$-form part of the solution has the same form as in the flat case.

The gravitational equations (\ref{2.4i}) are also satisfied by (\ref{5.pa})
with $\phi (t)$ given by (\ref{4.n}). It may be shown in the same way as it
was done in \cite{IS} that, due to the (anti-)self-duality condition, the
energy-momentum tensor satisfies
\beq\label{5.d}
     	T[F,g] _{ij} = 0 \qquad \Rightarrow \qquad T[F,g]_i^{~j} = 0,
\eeq
for all $i,j = 1, \dots, n$, i.e., the form field contributes as dust matter.

\section{Generalization to a set of extra Ricci-flat spaces}

Here we give a further generalization of the solution of the
previous section to the case when the manifold (\ref{3.10g}) is replaced by
\beq\label{3.10gi}
       M = (t_{-}, t_{+})  \times N \times N_1 \times \ldots N_k,
\eeq
where $N_r$ are Ricci-flat manifolds with the metric $h^r$ of
dimension $d_r$. The solution reads
\bearr			\label{4A.pai}
     ds^2 = \exp \left[ \frac{4m f(t)}{K(2-D)} \right]
      \biggl( - \e^{2c^0 t + 2 \bar c^0}  dt^2
\nnn  \qquad
      + \e^{f(t)/K} h_{ij}(y)dy^i dy^j
     	+ \sum_{r=1}^{k}  \e^{2c^r t+2 \bar{c}^r} ds^2_r  \biggr),
\\  \lal \label{4A.pci}
     F = \e^{2 f(t)} dt \wedge Q,
\ear
where $ds^2_r = h^r_{m_r n_r}(z_r) dz_r^{m_r} dz_r^{n_r}$ is the line element
corresponding to the metric $h^r$, the constants $c^0$, $\bar c^0$ are
given by
\beq\label{4A.ci}
     c^0  = \sum_{r=1}^k d_r c^r, \cm
     \bar c^0  =  \sum_{r=1}^k d_r \bar c^r
\eeq
and the function  $f(t)$ is again given by
\beq\label{4A.i}
        f(t) = - \ln \left[|z(t) ||K Q^2|^{1/2} \right]
\eeq
with
\bear\label{4A.ja}  \nq
     z(t) \eql \frac{1}{\sqrt{C}} \sinh \left[ (t-t_0) \sqrt{C} \right],
            K<0 , ~C>0;
\\                        \label{4A.jb}
	\eql\frac{1}{\sqrt{-C}}\sin\left[(t-t_0)\sqrt{-C}\right],
        	K<0,~C<0;
\\                                                  \label{4A.jc}
     \eql t-t_0, \qquad  \qquad A<0, ~C=0;
\\                                                  \label{4A.jd}
     \eql \frac{1}{\sqrt{C}} \cosh \left[ (t-t_0) \sqrt{C} \right],
     		 K>0 , ~C>0.
\ear
Here $D=4m+1+\sum_{r=1}^k d_r$ and
\beq						\label{4A.hbi}
     	K  =  m + \frac{4m^2}{2-D} \neq 0
\eeq
and the integration constants obey the relation
 \bear\label{4A.hai}
     C K^{-1} +  \sum_{r = 1}^{k} (c^r)^2 d_r
     - \biggl(\sum_{r = 1}^{k} c^r d_r\biggr)^2 = 0 .
\ear
This solution may be readily verified by substitution into the field
equations.

As a final remark, the metric (\ref{4A.pai}) may also be obtained from the
exact cosmological solutions for a one-component anisotropic (``perfect``)
fluid \cite{IM94,GIM}  when the pressure in the subspace $N$ is $p_0 = 0$
(``dust'') and the pressures in the extra subspaces $N_i$ are $p_i = \rho$
(``stiff'' matter), where $\rho$ is the energy density.  A systematic
derivation of this (and more general) solution will be given in a separate
publication.

\section{Summary and conclusions}

We have considered a $D =(n+1)$-dimensional cosmological model with an
antisymmetric $(p+2)$-form. We have generalized the composite electric
$S$-brane solutions from \cite{IMS} for $D = 4m+1 = 5, 9, 13, ...$ and $p =
2m-1 = 1, 3, 5, ...$ to the case when the $Q$-form of rank $2m$ is defined
on an $n$-dimensional oriented Ricci-flat space $N$ of Euclidean signature.
Here the form $Q$ is an arbitrary (anti-)self-dual parallel (i.e.,
covariantly constant) $2m$-form on $N$ with $Q^2 > 0$.  For the flat case,
when $N = \mathbb{R}^{4m}$ \cite{IMS}, the components of this form in
canonical coordinates are proportional to charge densities of electric
$p$-branes.

We have also found generalizations of the solutions to the case when a chain
of extra Ricci-flat factor-spaces is added.

\Acknow
 {The work of V.D.I. was supported in part by a grant of College of Science
 and Mathematics of California State University (Fresno), by DFG grant
 No. 436 RUS 113/807/0-1 and by the Russian Basic Research Foundation grant
 No. 05-02-17478. V.D.I. thanks the colleagues from the Physical Department
 of the California State University for hospitality during his
 visit in November-December 2005.}

  \end{document}